\documentclass{article}

\usepackage{arxiv}

\usepackage[utf8]{inputenc} 
\usepackage[T1]{fontenc}    
\usepackage{hyperref}       
\usepackage{url}            
\usepackage{booktabs}       
\usepackage{amsfonts}       
\usepackage{amsmath}        
\usepackage{nicefrac}       
\usepackage{microtype}      
\usepackage{lipsum}
\usepackage{graphicx}
\usepackage{amsbsy}
\usepackage{mathrsfs}
\DeclareMathAlphabet{\mathscrbf}{OMS}{mdugm}{b}{n}
\graphicspath{ {./figures/} }

\title{VARX Granger Analysis: Modeling, Inference, and Applications}
\shorttitle{VARX Granger Analysis}

\author{
Lucas C. Parra \\
  Biomedical Engineering\\
  City College of New York\\
  \texttt{parra@ccny.cuny.edu} \\
   \And
Aimar Silvan Ortubay\\
  Biomedical Engineering\\
  City College of New York\\
  \texttt{asilvanortubay@ccny.cuny.edu} \\
   \And
Maximilian Nentwich\\
   The Feinstein Institutes for Medical Research\\
   Northwell Health\\
   \texttt{max.nentwich@gmail.com}\\
   \And
Jens Madsen\\
     Biomedical Engineering\\
  City College of New York\\  \texttt{jmadsen@ccny.cuny.edu} \\
   \And
Behtash Babadi \\
  Electrical and Computer Engineering\\
  University of Maryland\\
  \texttt{behtash@umd.edu}
}

\begin{document}
\maketitle
\begin{abstract}
Complex systems, such as brains, markets, and societies, exhibit internal dynamics influenced by external factors. Disentangling delayed external effects from internal dynamics within these systems is often challenging. We propose using a Vector Autoregressive model with eXogenous input (VARX) to capture delayed interactions between internal and external variables. While this model aligns with Granger's statistical formalism for testing "causal relations", the connection between the two is not widely understood. Here, we bridge this gap by providing fundamental equations, user-friendly code, and demonstrations using simulated and real-world data from neuroscience, physiology, sociology, and economics. Our examples illustrate how the model avoids spurious correlation by factoring out external influences from internal dynamics, leading to more parsimonious explanations of the systems. We also provide methods for enhancing model efficiency, such as L2 regularization for limited data and basis functions to cope with extended delays. Additionally, we analyze model performance under various scenarios where model assumptions are violated. MATLAB, Python, and R code are provided for easy adoption: \url{https://github.com/lcparra/varx}
\end{abstract}

\section{Introduction}

Analyzing signals generated by real-world dynamical systems such as neural activity in the brain, physiological signals in the body, or trends in society and the economy is a key component of scientific discovery. These systems all involve endogenous variables, which are internal variables that develop and interact with each other over time. Additionally, these systems are influenced by exogenous variables, which are external factors that serve as drivers of the endogenous dynamics (for instance, a visual stimulus to the brain, or fiscal stimulus to the economy). It is often not clear how to separate external drive from the internal dynamics.  

A standard modeling approach to capture effects between dynamic variables is to determine if one variable can be predicted from another. Conventionally one focuses on linear prediction, which often captures the dominant relationships. When considering time delays, this translates to finding a linear filter that best predicts the next time point based on the preceding signal. When predicting a variable from its own past, this filter is referred to as an auto-regressive (AR) model. In scenarios with multiple endogenous variables, a vectorial auto-regressive (VAR) model is employed, characterized by multiple filters between all the variables. Finally, when some variables represent exogenous inputs, the corresponding model is known as a VARX model \cite{ljung1999system}.
 
To determine if the temporal predictions capture real statistical effects, Clive Granger proposed a method for computing statistical significance in his seminal work on causality \cite{granger1969investigating}. He essentially asks whether the quality of the prediction is significantly improved when a variable is added to the model. The statistical tests determines, specifically, whether the linear prediction of a variable $y(t)$ from its past can be improved by adding the history of another variable $x(t)$ as predictor. If such an improvement is observed, we say that $x$ has an "effect" on $y$. Granger referred to this as "causal relation". The basic idea had been suggested earlier by Wiener \cite{wiener1956theory},  and it is sometimes referred to as "Wiener-Granger Causality" \cite{bressler2011wiener}. In this work we avoid calling an effect "causal", due to several well-known limitations to this interpretation, which we will dis.

The statistical formalism developed by Granger can be applied naturally to determine the significance of the effects in VAR models \cite{shojaie2022granger}. The Granger analysis for VAR models has been useful to neuroscientists, economists, and sociologists because it allows one to quantify the strength and direction of effects in interactive dynamical systems, such as brains \cite{bressler2011wiener,seth2015granger}, markets \cite{qin2011rise}, and societies \cite{freeman1983granger}. However, researchers typically ignore exogenous variables during Granger analysis, for a lack of available tools to do so. While Granger and later Geweke allow for exogenous variables \cite{geweke1984measures}, these are only included to remove potential instantaneous confounds. In contrast, we treat exogenous variables in their own right by applying the Granger formalism to VARX models. By doing this, we are capturing the lagged effects of the exogenous input, and separating that from the internal dynamics. 

We begin by summarizing the basic equations required to estimate VARX model parameters, effect size, and statistical significance. Then, we demonstrate the validity of this approach using simulated data. Subsequently, we showcase examples that apply this formalism to neural signals, highlighting the key differences between VARX models and "temporal response functions" \cite{Crosse2016-lp} commonly used in neuroscience. We will also present examples using physiological, sociological, and economic data. Finally, we conclude with a discussion on the specifics of code implementation and some caveats regarding the interpretation of model results. 

An appendix will explore methods (L2 regularization and basis functions) to handle high-dimensional datasets and longer prediction filters. These methods effectively reduce the number of parameters. Notably, we will present the first derivation of de-biased estimates of the test statistic for L2 regularization. We will also illustrate instances where interpreting the Granger formalism as "causal" can be misleading, such as in cases involving missing variables or colliders.  

\section{Methods}

A way to understand the VARX model is to imagine pellets dropping into a pond. The pellets are like an external input (exogenous), but the ripples they create are governed by the water's own internal dynamics (endogenous). These ripples can also be influenced by unpredictable gusts of wind. Our goal is to distinguish between these external and internal factors by analyzing what we can observe – the pellets and the water's surface patterns – while minimizing the influence of the unseen wind.

\subsection{VARX model}

More formally, consider the vectorial “input” signal ${\bf x}(t)$ and the vectorial “output” signal ${\bf y}(t)$ of dimensions $d_x$ and $d_y$ respectively, with both assumed to be observable (lower-case bold characters represent vectors). In the case of brain activity, the input may be multiple features of a continuous natural stimulus, say luminance and sound volume of a movie. The output could be neural activity recorded at multiple locations in the brain (see Section~\ref{section:brain}). In the case of macroeconomic variables, the input could the government spending and the endogenous variables could be various indicators of economic activity of a nation (see Section~\ref{section:economy}). The simplest model we can envision is one where the current signal  ${\bf y}(t)$ can be predicted linearly from the input ${\bf x}(t)$ and also linearly from the preceding output ${\bf y}(t-1)$.  
\begin{equation}
{\bf y}(t) = {\bf A}*{\bf y}(t-1) + {\bf B}*{\bf x}(t) + {\bf e}(t)	.
\label{varx}
\end{equation}
Here, ${\bf A}$ and ${\bf B}$ are filter matrices of dimensions $[d_y,d_y]$ and $[d_y,d_x]$, with filters of length $n_a$ and $n_b$, respectively. The additive term ${\bf e}(t)$ represents an unobserved "innovation" that introduces an error in the prediction. In linear systems it is called innovation because it injects novelty into the recurrent dynamic. We refer to ${\bf y}(t)$ as endogenous variables, as they are influenced by one another including their own history (through the diagonal terms in ${\bf A}$, and to ${\bf a}(t)$ as exogenous variable as they are fixed and given, and not influenced by the endogenous variables.   

In Eq. (\ref{varx}) we have used a compact formulation of a multi-input multi-output convolution, which for the auto-regressive filters ${\bf A}$ and moving average filters ${\bf B}$ reads: 
\begin{eqnarray}
{\bf A}*{\bf y}(t-1) &=& \sum_{l=1}^{n_a} {\bf A}(l)\cdot{\bf y}(t-l) ,
\label{A-convolution}
\\
{\bf B}*{\bf x}(t) &=& \sum_{l=0}^{n_b-1} {\bf B}(l) \cdot {\bf x}(t-l)  .
\end{eqnarray}

\begin{figure}[ht]
\begin{center}
\includegraphics[trim={0 0 10.5in .7in},clip,width=.7\columnwidth]{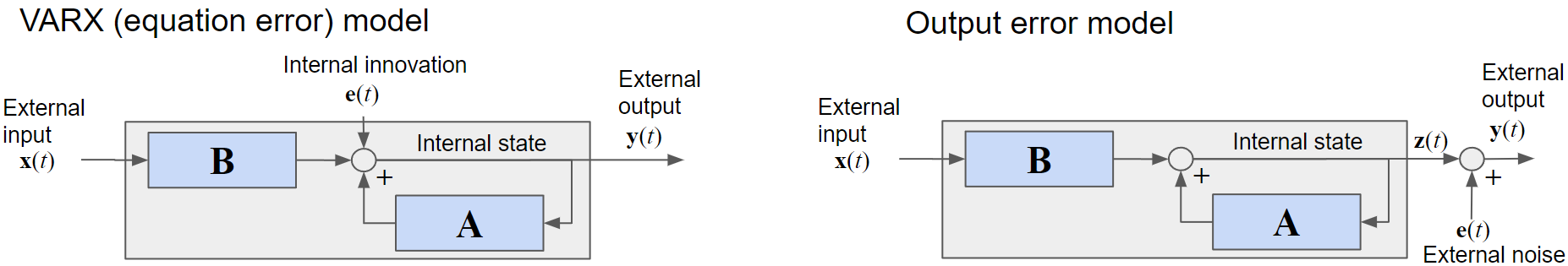}
\end{center}
\caption{VARX model: The gray box represents the overall system response $\mathcal{H}$.}
\label{schematic}
\end{figure}

\subsection{Total system response}

Note that the total response of the dynamical system (the impulse response) can be written in the $Z$ or Fourier domains simply as
\begin{equation}
\mathcal{H} = \mathcal{Y}  \mathcal{X}^{-1} = (\mathcal{I} - \mathcal{A})^{-1} \mathcal{B}.
\label{system-response}
\end{equation} 
In this view, what we are proposing is to model the total system response as a combination of a Moving Average (MA) filtering ${\bf B}$ and Auto-Regressive (AR) filtering $({\bf I}-{\bf A})^{-1}$. In the time domain, the total system response ${\bf H}(t)$ can simply be computed by passing impulses in each input variable through the system, while setting the error/innovation to zero, ${\bf e}(t)=0$.

\subsection{System identification}

Given observed ${\bf x}(t) and {\bf y}(t)$ one can estimate the parameters ${\bf A}$ and ${\bf B}$ by minimizing the mean of square error:
\begin{equation}
{\boldsymbol\sigma}^2 = \frac{1}{T} \sum_{t=1}^T {\bf e}^2(t). 
\end{equation}
This identification criterion is equivalent to Maximum Likelihood estimation, assuming the innovation ${\bf e}(t)$ is normal distributed and uncorrelated (spherical and white). For zero-mean signals this is also the variance, hence the conventional symbol ${\boldsymbol\sigma}^2$, which is vectorial here as it is computed and minimized for each dimension in ${\bf y}$ individually. For the VARX model, the linear predictors ${\bf y}(t-1)$ and ${\bf x}(t)$ are observable (this will not be true for the Output Error model discussed in the Appendix Section~\ref{section:output-error}), and parameter estimation results in a simple linear least-squares problem with a well-established closed form solution. Equation (\ref{varx}) can be rewritten as
\begin{equation}\label{eq:linear-model}
{\bf Y} = {\bf X}\cdot \bf H + {\bf E},						    
\end{equation}
where ${\bf X}$ is a block-Toeplitz matrix of the predictors, including ${\bf y}(t-1)$ and ${\bf x}(t)$. This matrix has dimensions $[T, N]$, where $N$ is the total number of free parameters for each predicted dimension in $y(t)$. In this case $N=d_y n_a + d_x n_b$. ${\bf Y}$ is the output signal ${\bf y}(t)$ arranged as a matrix of dimensions $[T,d_y]$, and ${\bf H} = [{\bf A}, {\bf B}]^\top$ is a matrix of dimensions $[d_y, N]$ combining the AR and MA filters. The least-squares estimate is then simply: 
\begin{equation}
\hat{\bf H} = {\bf R}_{xx}^{-1} {\bf R}_{xy} .
\end{equation}
Matrices ${\bf R}_{xx}$ and ${\bf R}_{xy}$ are block-toeplitz capturing cross-correlations:
\begin{eqnarray}
{\bf R}_{xx} = {\bf X}^\top \cdot {\bf X},
\\					
{\bf R}_{xy} = {\bf X}^\top \cdot {\bf Y}.					
\end{eqnarray}
The estimated output is:
\begin{equation}
    \hat{\bf Y} = {\bf X} * \hat{\bf H} .
    \label{Y=X*H}
\end{equation}
The residual errors of this model prediction for each output channel are the diagonal elements of the correlation matrix for the errors:
\begin{eqnarray}
{\bf R}_{ee} &=& ({\bf Y}-\hat{\bf Y})^\top\cdot({\bf Y}-\hat{\bf Y}) =  \hat{\bf H}^\top \cdot {\bf R}_{xx} \cdot \hat{\bf H} - 2 \hat{\bf H}^\top \cdot {\bf R}_{xy} +{\bf R}_{yy}, \\
\boldsymbol\sigma^2 &=& \frac{1}{T}\mathbf{diag}({\bf R}_{ee}).
\end{eqnarray}

In Section \ref{section:L2-regularization} we will discuss how these expressions change when we employ L2 regularization to mitigate overfitting. In Section~\ref{section:basis-functions} we will in addition extend the approach to include basis functions to represent filters ${\bf B}$ so as to reduce the number of free parameters, again with the goal of reducing overfitting. 

Note that Equations (\ref{A-convolution})-(\ref{Y=X*H}) are identical to modeling the total system response as a single multivariate MA filter, whereby matrix ${\bf X}$ only contains the input ${\bf x}(t)$, with $N= d_x n_b$, and the impulse response ${\bf H}={\bf B}$. In the neuroscience literature this MA model is referred to as a “multivariate temporal response function” (mTRF).

\subsection{Granger formalism} 
To establish if any of the channels in filters ${\bf A}$ or ${\bf B}$ significantly improve predictions -- i.e. have an "effect" -- one can use a likelihood-ratio test \cite{Geweke1982-am}. In this formalism, one uses deviance as the test statistics to quantify the contribution of a given predictor in ${\bf X}$ for each output in ${\bf Y}$. The approach consists of estimating filter parameters ${\bf H}$ with all predictors included in ${\bf X}$, which is referred to as the “full” model, and then again with one of the predictors removed, which is referred to as the “reduced” model. We compute the resulting square error $\hat{\boldsymbol\sigma}_f^2$ and $\hat{\boldsymbol\sigma}_r^2$, for the full and reduced models, and obtain the \emph{deviance} between the two models as test statistics (there is one deviance value for each dimension in ${\bf y}$): 
\begin{equation}
\mathscrbf{D} := T  \log \left (\hat{\boldsymbol\sigma}^2_r / \hat{\boldsymbol\sigma}^2_f \right),    
\end{equation}
where the division of the two variance vectors and the log operator are interpreted element-wise. 
For normal, independent, and identically distributed error, the vector $\mathscrbf{D}$ contains the log-likelihood ratios (times a factor of 2), with each element following a chi-square distribution \cite{Davidson1961-dr}. Notice that the test statistics vector $\mathscrbf{D}$ is formed by computing the log-likelihood ratio for each output dimension, and for each predictor dimension that is removed in the reduced model. Thus, one can estimate the statistical significance of each channel in ${\bf A}$ and ${\bf B}$ by computing the full model once, and then removing each predictor variable individually from the full model. The statistical significance for a non-zero contribution from a particular predictor to a particular output is then given by an element of the “p-value” vector computed with the corresponding deviance vector:  
\begin{equation}
\mathbf{p} = \mathbf{1} - F (\mathscrbf{D}, n),
\label{p-value}
\end{equation}
Here $F$ is the cumulative distribution function for the chi-square distribution and $n$ is the number of parameters that were removed in the reduced model, i.e $n_a$ or $n_b$ depending if an element of ${\bf y}(t-1)$ or ${\bf x}(t)$ was removed. The operation of $F$ on a vector is interpreted element-wise. 
 
\subsection{Effect size}

Note that deviance increases linearly with $T$, that is, the statistical evidence increases with the length of the signals, and can therefore not serve as effect size. A traditional definition of effect size in the context of reduced and full linear models is the coefficient of determination, or generalized R-square \cite{Magee1990-fx}:
\begin{equation}
R^2 := \mathbf{1} - \exp\left( - \mathscrbf{D}/T\right),
\label{effect-size}
\end{equation}
where the exponential of a vector is interpreted element-wise. 

\section{Results}

Details on all results provided next can be found in the accompanying Matlab code repository \url{https://github.com/lcparra/varx}. Code is also provided in Python and R. 

\subsection{Test of model estimation on known model}

\begin{figure}[ht]
\begin{center}
\includegraphics[width=1\columnwidth]{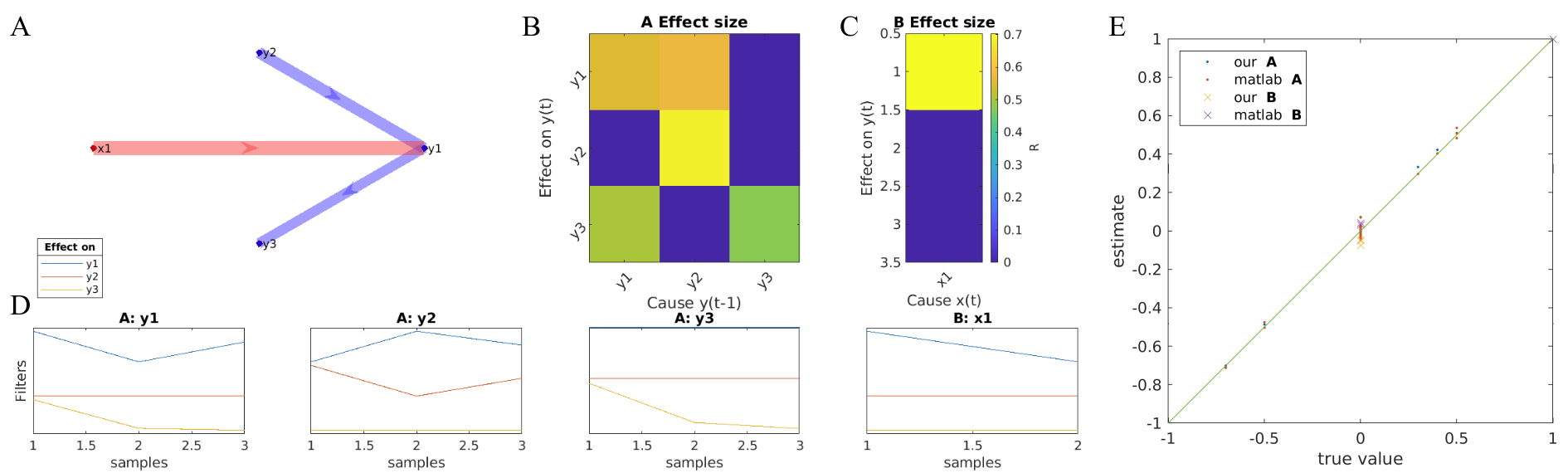}
\end{center}
\caption{Comparison of estimated parameters to true parameters in a simple toy example. Here $d_y=3, d_x=1, n_a=3, n_b=1$. A: Graph shows the effect sizes $R$ indicating the structure and direction of effects (red for exogenous effect ${\bf B}$, and blue for endogenous effects ${\bf A}$. B: Effect sizes $R$ now shown as a connectivity matrices. C: Estimated filters ${\bf A}$ and ${\bf B}$. D: comparison of true and estimated parameters, and comparison for results from Matlab's econometric toolbox (we used $n_b=1$ to satisfy the limitation of this toolbox). Signals were simulated for $T=1000$ time steps with $x(t) \sim \mathcal{N}(0, 1) \text{ i.i.d.}$, ${\bf e}(t)$. We used no L2 regularization and set $n_a=3, n_b=1$ for the estimation.  
}
\label{known-model}
\end{figure}

To validate the estimation algorithm and code, we simulated a simple VARX model with two outputs and one input ($d_y=2, d_x=1$). The algorithm correctly recovers the AR and MA parameters (Fig.\ref{known-model}).  VARX model estimation is available as part of the econometric toolbox in Matlab, but is limited to instantaneous input $n_b=1$, i.e. no filtering of the input. When limiting the simulation to this case, the algorithms obtain similar results. Small variations are expected based on how the initial boundary conditions are handled and numerical differences. In our implementation we omit from the estimate all samples that do not have a valid history. The code handles missing values (NaN) in the same way. 

\subsection{Validation of p-values}
\begin{figure}[ht]
\begin{center}
\includegraphics[width=0.8\columnwidth]{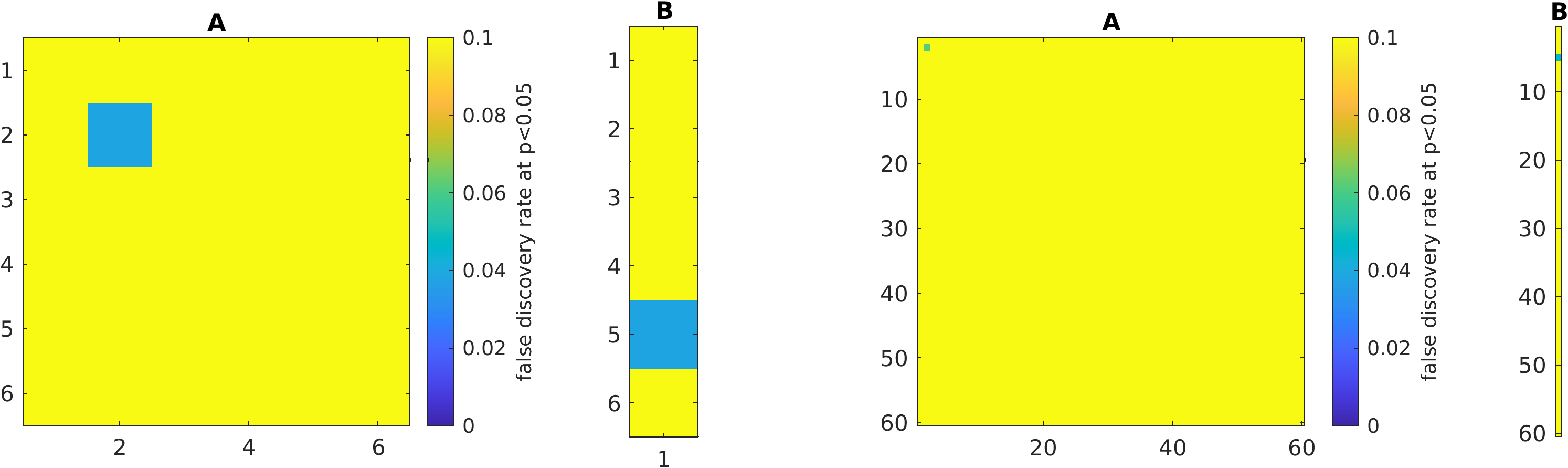}
\end{center}
\caption{Numerical validation of $p$-values in both a smaller and a larger model ($d_y=6$ and $d_y=60$). Significance is set of $p<0.05$, so we expect a false discovery rate of 0.05. Simulation here used with $d_x = 1, n_a=2, n_b=2, T=1000$. Filter coefficients for ${\bf B}$ were selected at random from unit variance normal, and ${\bf A}$ values were set to be $\pm 0.05$ with sign selected at random (this insured stable recursion in practice). Only two channels are set to zero $A(:,2,2)=0, B(:,5,1)=0$. For these two, the false discovery is is correctly estimated at approximately $0.05$ (green).   
}
\label{small-large-model}
\end{figure}

To validate the accuracy of the $p$-value estimation, we simulated a VARX model with all channels assigned random non-zero values, except for one channel in matrix ${\bf A}$ and one in matrix ${\bf B}$, which were set to zero. We did this with a small and large simulated dataset, generated with normal i.i.d. innovations ${\bf e}(t)$. We repeat the simulation 1000 times and determine how many times the zero channels report a $p<0.05$, i.e. we numerically estimate the false discovery rate. We find a false discovery rate of approximately 0.05 for the null-channel, suggesting that $p$-values are correctly estimated (Fig.~\ref{small-large-model}). For all others, the chance of detecting the non-zero effect is 1, i.e. a perfect true positive rate.  

\subsection{Example: Brain signals in human}
\label{section:brain}

A key advantage of VARX models lies in their ability to factorize the overall system response into AR and MA components, as shown in Eq.~(\ref{system-response}). This separates the influence of endogenous variables into an initial response followed by ongoing reverberations within the dynamical system. To illustrate this, we analyze intracranial electroencephalography (iEEG) recordings from a 49-year-old male patient watching movie clips (data from \cite{Nentwich2023-dz}). We focus on 50 electrodes in visual brain regions (occipital cortex, fusiform face area, and parahippocampal cortex). We use the neural data from \cite{Nentwich2023-dz}, and use the same preprocessing to extract high-frequency broadband activity (BHA, 70-150Hz, downsampled to 60Hz), often considered a marker of local neuronal firing. The exogenous "input" is a pulse train indicating fixation starts (moments of new visual input). 

\begin{figure}[ht]
\begin{center}
\includegraphics[width=1\columnwidth]{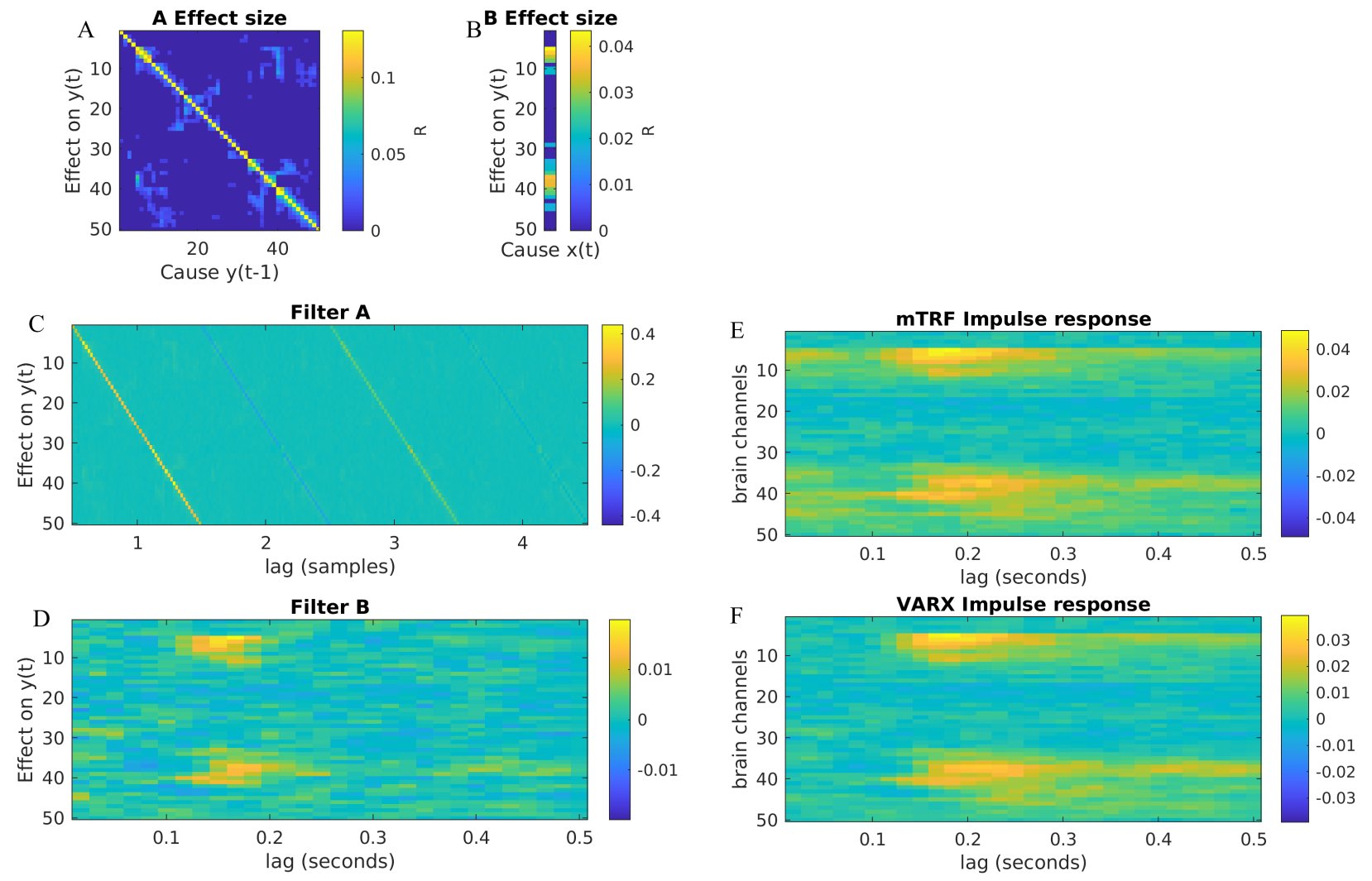}
\end{center}
\caption{Example of intracranial recording in humans: a VARX model was fitted to broadband high-frequency activity during free viewing of 7 videos recorded from 50 ($d_y=50$) electrodes. A total of 43.6 minutes of data was used at a sampling rate of 60 Hz ($T=156,955$) from a single patient. (A) Effect size $R$ for the recurrent connectivity ${\bf A}$ between recording electrodes -- in the language of neuroscience, this could be called "functional connectivity". (B) Effect size $R$ of fixation onset as exogenous variable on different electrodes. (C) ${\bf A}$ filter coefficients ($n_a=4$). (D) ${\bf B}$ filter coefficients $n_b=30, \underline{n}=20$. (E) System response estimated as a multivariate MA filter  -- in the language of neuroscience, this is the multivariate "temporal response function" (mTRF). (F) System response resulting from the VARX model estimate (Eq.~\ref{system-response}). Data from \cite{Nentwich2023-dz}.} 
\label{intracranial}
\end{figure}

The first observation is the diagonal terms of ${\bf A}$ dominate (Fig.~\ref{intracranial}A), with oscillating parameters (indicating a high-pass filter) (Fig.~\ref{intracranial}C). What is most evident is that the ${\bf B}$ (Fig.~\ref{intracranial}D) response is shorter than the total system response ${\bf H}$ (Fig.~\ref{intracranial}F). This suggests that the VARX model decomposes the total response into a fast response followed by a prolonged response due to the recurrence in the brain network. Estimating ${\bf H}$ directly (following \cite{Crosse2016-lp}, Fig.~\ref{intracranial}E), or using the VARX model, i.e. as ${\bf H} = (1-{\bf A})^{-1} {\bf B}$ (Fig. 4F), we see that the two are very similar. The factorization of the total system response in the VARX model, Eq.~(\ref{system-response}), thus appears to be a good approximation of the direct estimate as a purely MA system response. 

We employed basis functions to represent long delay filters ${\bf B}$ of length $n_b$ efficiently with fewer parameters, namely $\underline{n} < n_b$ (see Section \ref{section:basis-functions} for details).

\subsection{Example: Physiological signals in human}

\begin{figure}[ht]
\begin{center}
\includegraphics[width=0.8\columnwidth,trim={0 3.25cm 0 3cm},clip]{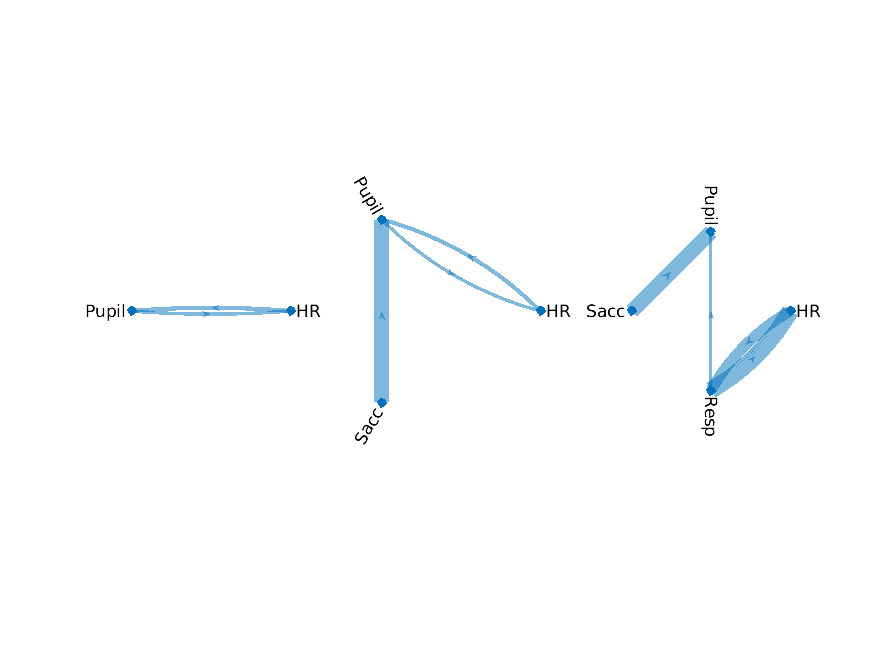}
\end{center}
\caption{Example of physiological signals in human. This data was collected while study participants carried out a rhythmic breading task. In this case there was no exogenous stimulus, so we only fit a VAR model. Links in ${\bf A}$ are show if $p<0.001$. Here we had 26 minutes of data from compiled across multiple subjects samples at 25Hz ($T=26*60*25$). Data from \cite{madsen2024bidirectional}. 
}
\label{brain-body}
\end{figure}

Human physiology is a dynamic system with multiple dependent signals like respiration, heart rate, pupil response, and brain activity. Analyzing data from \cite{madsen2024bidirectional}, which demonstrated bidirectional effects between several signal modalities, our VARX analysis indicates potential directional effects among these physiological variables (Fig.~\ref{brain-body}).

Pupil size and heart rate were measured in the experiment as metrics of physiological arousal. As variables are added to the VARX model the connectivity structure is typically preserved. In this specific example, using a controlled breathing task, we initially observe a bidirectional link between pupil size and heart rate (Fig.~\ref{brain-body}A), however this disappears once respiration is taken into account (Fig.~\ref{brain-body}B). Instead, this link is explained by an effect of respiration on pupil size, together with the well-established bidirectional link between respiration and heart rate that is recovered in this data. Saccades, which are short, rapid eye movements, also have a well-established effect on pupil size identified in this study (Fig.~\ref{brain-body}C).

In general, adding variables can remove links -- if the new common-cause variable provides an explanation for a spurious link. Adding variables can also add links -- if the addition is a "collider". This is well established for i.i.d samples \cite{Pearl2013-ce}, and is no different for temporally correlated time sequence data. We will demonstrate this further using simulated data in Section~\ref{system-response}.  

\subsection{Example: Union participation in the US}

\begin{figure}[ht]
\begin{center}
\includegraphics[width=0.8\columnwidth]{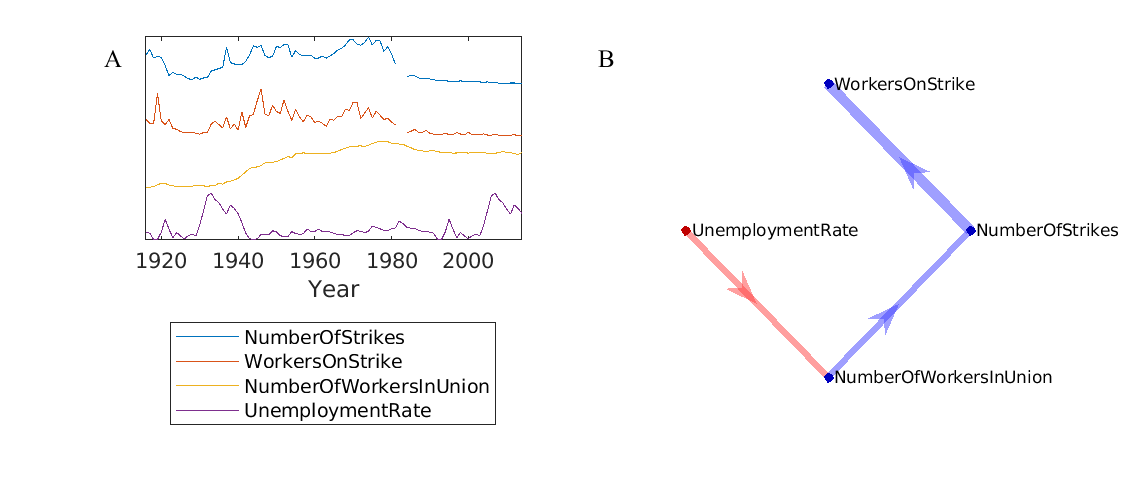}
\end{center}
\caption{Example on union participation and strikes: (A) Historical data from the US. We treated the unemployment rate as an exogenous input in the VARX model, and the others as endogenous variables. (B) Significant effects in ${\bf A}$ and ${\bf B}$  are indicated in blue and red, respectively ($p<0.05$,  $n_a=n_b=3, T=195$). Note missing data around 1980, which was omitted during the estimation, including a 3-year history.}
\label{union-history}
\end{figure}

Here, we present an analysis from the field of sociology. We examine the history of workers' union membership and its relationship to strikes (Fig.~\ref{union-history}A). We hypothesize that strikes increase union membership in subsequent years. We assume the unemployment rate is unaffected by union variables, so it is modeled as an exogenous input. In contrast, the number of unionized workers, the number of workers on strike, and the number of strikes can all potentially influence each other.  VARX Granger analysis (Fig.~\ref{union-history}B) suggests that unemployment affects unionization, which in turn affects number of strikes, which obviously affects the number of workers on strike. These results depended on the choice of hyper-parameters $n_a, n_b, \lambda$. Only the effect of NumberOfStrikes $\rightarrow$ WorkersOnStrike was robust to parameter choice. What did not robustly emerge from this data is evidence for the initial hypothesis that strikes lead to an increase in union membership.

\subsection{Example: Macroeconomic dynamic in the US}
\label{section:economy}

\begin{figure}[ht]
\begin{center}
\includegraphics[width=0.8\columnwidth]{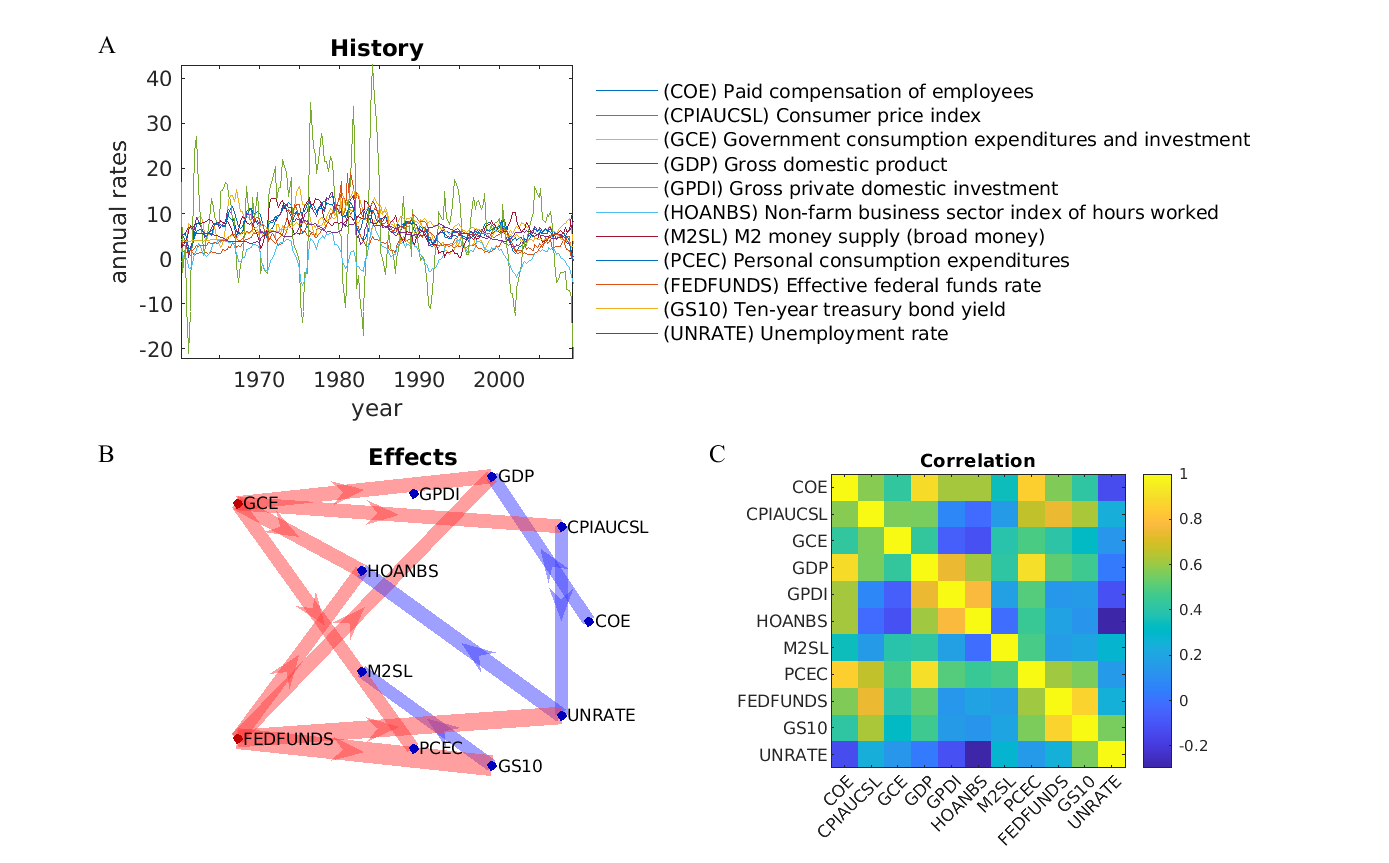}
\end{center}
\caption{Example on US macroeconomic data: (A) Historical data from the US measured every quarter. None-rate variables (1-8) have been converted into annual percentage rates of change. (B) Significant effects in ${\bf A}$ and ${\bf B}$  are indicated in blue and red, respectively ($p<0.001, T=195$). We are taking 12-18 months history into account ($n_a=4, n_b=6$). (C) Pearson correlation of all variables.}
\label{US-Economy}
\end{figure}

As a final example, we demonstrate the model on a dataset from the U.S. Federal Reserve encompassing fiscal, monetary, and labor factors, spanning quarters from 1959 to 2009. Here we have converted all gross numbers into annual percentage rates. This removes the exponential growth resulting from predominantly positive rates (Fig.~\ref{US-Economy}A), which leads to trivial correlations and non-stationarity (sometimes referred to as unit-root signals). To determine the effect of the government on the economic variables, we examined the impact of government spending (GCE) and federal funds rate (FEDFUND). Government spending itself is a function of economic conditions, such as unemployment benefits, which are automatically linked to unemployment, while a rise in GDP increases tax revenue, which typically leads to increased government spending. Nevertheless, by treating GCE and FEDFUND as exogenous variables, we are asking what effects these government policies have on the economy, if they were controlled independently. Before we discuss the results (Fig.~\ref{US-Economy}B), it is important to note that the specific effects strongly depend on the choice of variables (gross numbers vs annual rates, endogenous vs exogenous) and parameters (independent of hyper-parameters $n_a, n_b, \lambda$). However, a robust finding is the direct effect of government spending on gross domestic product (GDP), inflation (CPIAUCSL) and personal spending (PCEC). Rate policy affects the unemployment rate (UNRATE) independently of government spending. Despite the sensitivity to parameters, the model identifies sensible relationships and demonstrates that many variables remain independent despite a dense correlation structure (Fig.~\ref{US-Economy}C).   

In the Appendix we present results related to L2 regularization (Section~\ref{section:L2-regularization} and \ref{section:validation-debiased-deviance}), including a derivation for the de-biased deviance for L2 regularization, that is not yet in the literature (Section~\ref{section:derivation-debiased-deviance}. We also extend the method to include basis functions and validate the p-value calculation on simulated data (Section~\ref{section:basis-functions}). We also present a number of scenarios where the model assumptions do not match the analytic model ((Section~\ref{section:confound-collider}, and~\ref{section:output-error}). These scenarios justify the use of the VAX model in some instances, and suggest caveats to the interpretation in others, e.g. when observations of important dynamic are missing. 

\section{Discussion}

Here we will first discuss the novel contributions of this work and compare our code to existing software tools. We then follow with a number caveats and methodological comments. 

\subsection*{Novel contributions}
Exogenous variables were incorporated into the Granger formalism as conditional dependence by Geweke \cite{geweke1984measures} and were already briefly discussed in the original work of Granger~\cite{granger1969investigating}. In practice, this "conditional causality" has been used to control for spurious correlations due to common causes. Code implementations of this idea \cite{guo2008partial,Barnett2014-rr} only use exogenous variables to remove confounds. In contrast, we propose to model the total system response as a combination of both exogenous effects and endogenous dynamics. In this view, the exogenous effects are not a nuisance, but an important component of the model to be estimated, encompassing multiple time delays. Although economists have employed VARX models to capture exogenous effects, the use of the Granger formalism to establish the effects of individual variables is not as widely used. Indeed, the correspondence of the "conditional Granger-causality" \cite{bressler2011wiener,ding2006granger} with VARX models is not well known. While the VARX model is common in statistics toolboxes, we are not aware of any implementation of the VARX model with the Granger-Geweke test to assess the effects. 

\subsubsection*{Toolboxes}
There are several software tools to estimate VARX models. We are not aware of one that provides the Granger-Geweke test for significance. The MVGC toolbox \cite{Barnett2014-rr} does not report results on exogenous variables, \url{https://github.com/SacklerCentre/MVGC1}. Implementations of VARX models in Matlab Econometrics Toolbox and in SAS software for instance, make significance statements for individual delays but do not allow for the exogenous variable, i.e. $nb=1$. We welcome feedback if there are other tools that should be mentioned here. Our code emphasized computational efficiency to handle large datasets, with comparatively long ${\bf B}$ filters. 

\subsection*{Equation error versus output error models} 
The estimation of model parameters for a VARX model has a closed-form solution, thus being much faster than finding parameters for an output error model, which requires iterative algorithms \cite{ljung1999system}. The gain in computational efficiency results from the assumption that $y(t)$ is observable. This may not be a good assumption in the case of brain signals measured across the skull, such as EEG/MEG where only a linear mixture, possibly with added noise, is observed. In that case, iterative algorithms are needed, but the Granger formalism can still be used with some effort \cite{Soleimani2022-bh}. In the VARX model, however, we do not need to assume that all internal activity is directly observable. Any unobserved activity is captured as innovation $e(t)$. We only need to be aware that any recurrent connectivity may be due to those unobserved common "causes". In particular symmetric effect sizes $R$ will be suggestive of such a missing variable. The role of the error is quite different in the two models. In the VARX (equation error) the error is an internal source of innovation driving the recurrent dynamic similar to the drive that comes from the input. The internal states are fully observable. In the output error model, the input entirely drives the system, and the error only affects the observations and is not injected into the dynamic. 

\subsection*{Comparison of VARX, MA, OE, and VAR models} 
VARX and output error (OE) models can be viewed as ways to break down a system's response into moving average (MA) and autoregressive (AR) components. Alternatively, the entire system response can be modeled as a pure MA filter, as demonstrated in Fig.~\ref{intracranial}E. In theory, incorporating an AR component allows for the representation of long impulse responses with fewer parameters, which is a practical advantage. However, the key difference lies in the error assumptions: MA and OE models assume errors at the output, while VARX models assume an internal innovation process with no error in the observations. Therefore, VARX models should not be considered mere input-output models, but rather as models of internal dynamics. It's worth noting that all variables can be included in a AR portion of the model, allowing the estimation process to determine if any variable acts is as an external input (i.e. that it does not depend on any other variable). For example, in the US macroeconomic model, arguably, government spending should have been included in the AR portion of the model, as it may depend on other variables. Including variables as exogenous serves to incorporate prior knowledge, like knowing that movie stimuli cannot be caused by brain activity. Additionally, it allows for counterfactual analysis, such as exploring the effects of independently controlled government spending. 

\subsubsection*{Sensitivity to parameters}
A caveat to all results above is that individual links can be sensitive to the model assumptions, namely, which variables are selected as endogenous (and can be affected by all others), and which variables are selected as exogenous (and cannot be affected). An example of that was the choice of the unemployment rate as exogenous to the dynamic of unions. The results can also depend on which endogenous variables are included, as we saw in the example of physiological signals. Results can also depend on the number of parameters $n_a$ and $n_b$ and regularization factor $\lambda$ (we saw this in examples with Unions and the US macroeconomic data). Further investigation on robustness to parameter choice is required for a clear interpretation of those results. Although we did demonstrate this here, these parameters could be established with cross-validation.   

\subsubsection*{Causality}
In Granger's original work \cite{granger1969investigating}, the error of the full and reduced model refers to one-dimensional signals where $y(t-1)$ is used in both cases and $x(t)$ is either used or omitted. If the error is significantly reduced by including $x(t)$ in the model, Granger argues that $x$ "causes" $y$. This interpretation is problematic for several reasons \cite{maziarz2015review}. As we saw, when common causes are not observed (either as external input or internal variables) they can generate spurious links \cite{lutkepohl1982non}. Bidirectional effects between two variables (e.g.~Fig.~\ref{US-Economy}) may be an indication of an underlying unobserved common cause. Similarly, including colliders can cause spurious links. All this is well explained by Pearl's approach to causal inference \cite{Pearl2013-ce}. Therefore one should not think of the Granger formalism as serious evidence for a causal graph without a well-justified prior graphical model \cite{pearl2009causality}. In particular for large dimensional datasets such as brain data, where we only observe a tiny fraction of all the variables, the risk of unobserved common causes is much too large to take the resulting graph seriously as a causal graph. Nevertheless, asymmetries in the ${\bf A}$ matrix can be seen as evidence of temporal precedence suggestive of an asymmetric "information flow". 

\subsubsection*{Non-stationarity}
Deviance makes a statistical judgment for the entire channel, not individual delays, as is common we simply treat each delay as a new predictor with its statistical test (this is the approach of the Matlab VARX). There are multiple methods under the umbrella of "Granger causality" that attempt to decide on how many tabs or which delays to use. By collapsing statistical evidence into a single test statistic, Deviance, this approach has greater statistical power. This is reflected in the linearity of $D$ with the number of samples $T$. The flip side is that this statistic is very sensitive to violations of its assumptions. For instance, it assumes that all $T$ samples of the innovation process are independent and identically distributed. The AR portion of the model assures that the linear-fit residual errors are uncorrelated in time, however, if there is any non-stationarity, this will no longer be the case. Therefore, non-stationarity will cause spurious correlations \cite{granger1974spurious}. In particular, any transient will cause larger deflection and correlation across samples. In particular, transient that affect several signals, say a common edge at the start or end of the signal may appear to behave like a common drive with high amplitude that results in a spurious link. Therefore, in the present approach, one has to treat edges and transients with utmost care to avoid spurious links. 

Some have argued that issues with non-linearity and non-stationarity can be addressed \cite{Barnett2015-vp}. Barnett et al. proposed a State Space model that can cope to some degree with missing variables, does not need to compute a reduced model, and can deal with non-linearity and non-stationarity \cite{Barnett2015-vp}. However, Stokes and Purdon showed that even the state-space Granger is not immune to confounding effects, non-stationarities, etc. The topic remains a matter of debate \cite{barnett2018misunderstandings}.

An alternative is to avoid using analytic expressions for the $p$-value, Eq. (\ref{p-value}), and instead use standard non-parametric statistics. For time series, the simplest is to randomly time-delay channels relative to one-another, potentially with a circular wraparound. All else in the model identification, i.e. estimates ${\bf A}$, ${\bf B}$ and effect size $R^2$ remain valid estimates of linear predictions even in the presence of non-stationarity and non-linearity. 

\subsubsection*{Stability}
A word about ${\bf A}$ is in order. The AR filter $1/(1-{\bf A})$ can be unstable. We have not implemented any mechanism for this vectorial AR filter to remain stable. Lack of stability only manifests in the systems when computing the overall systems response $\mathcal{\bf H}$, which is not necessary during estimation for the ${\bf B},{\bf A}$ nor the calculation of statistical significance of each path (contrary to the output error model, where the recurrence has to be run back in time to estimated gradients, risking issues of stability. We rarely encountered unstable AR estimates, and where we did, L2 regularization addressed the issues. But again, there is nothing in our formalism to ensure the stability of  $\mathcal{\bf H}$. 

\subsection*{1/f spectrum}
Another word about ${\bf A}$. The diagonal elements of ${\bf A}$ in practice will always be high-pass filters, as we saw in the example of intra-cranial recordings. We advise not to take individual delays in the diagonal terms literally. The reason for this is that the innovation process is assumed to be white (constant spectrum), whereas all natural signals tend to have a 1/f spectrum. As a result, $1/(1-{\bf A})$ has to have a 1/f spectrum, and ${\bf A}$ has to scale with f, i.e. be high-pass. In practice, we find that this is entirely accomplished by the diagonal elements of ${\bf A}$. But the caveat in principle applies also to the off-diagonal elements. Future work could consider a VARMAX model where the innovation is first filtered and then injected into the recurrent dynamic \cite{ljung1999system}. However, estimation of VARMAX model parameters is a non-convex optimization problem with similar complications to the output-error model.     

\section{Conclusion}

The predominant approach to modeling the effect of exogenous variables onto a dynamical system is to simply treat them as input and output of a vectorial MA filter (known as "temporal response function" in neuroscience, or simply "impulse response" in the linear systems literature). Unlike the VAR model, this is not commonly examined in the Granger formalism. Although Granger and Geweke both incorporate exogenous variables into the analysis formalism \cite{geweke1984measures}, the connection to the VARX model has not yet widely recognized.  We hope to have bridged this gap. While not incorrect, the simple MA approach fails to factor out the portion of the total system response is due to the internal dynamic, and separate that from the external drive. When relying only on VAR models, one fails to exploit the prior knowledge that some variables are independent of the internal dynamic. In summary, different models vary in their assumptions about how to break down the system's overall response. When estimating with the VARX model, we manage to uniquely factor the overall response into external drive versus internal dynamics.

\section{Acknowledgement}

We would like to thank Vera E. Parra for posing the question on union participation and compiling the corresponding datasets. We also thank Jacek Dmochowski for conversations about the overall system response of the VARX model. We thank Alain de Cheveigne for providing feedback on an earlier version of this manuscript. We thanks Wim Vijverberg for general advice on the analysis of macroeconomic data. 

\section{Appendix: Further validation and extensions}

\subsection{Debiased Deviance for L2 Regularization}
\label{section:L2-regularization}

To avoid overtraining on small sample sizes, i.e. where $T$ is not much larger than $N$, we decided to use an L2 penalty, with Tikhonov regularization. The advantage over other forms of regularization, such as L1 \cite{Das2023-fk,Soleimani2022-bh} or a state space model \cite{Barnett2014-rr,Barnett2014-rr} is computational efficiency thanks to the closed-form solution:
\begin{eqnarray}
\hat{\bf H} = ({\bf R}_{xx} + \gamma {\bf\Gamma}) ^{-1} {\bf R}_{xy},
\end{eqnarray}
where we selected ${\bf \Gamma}= {\rm diag}({\bf R}_{xx})$ so that all variables are regularized equally regardless of their scale. The choice of $\gamma$ is discussed in the results section. This regularization introduces (purposefully) a bias in the estimate, and the deviance estimate has to be corrected to account for this bias \cite{Sheikhattar2018-dc}. The term that corrects the log-likelihood in case of L2 regularization is (derived in Section~\ref{section:derivation-debiased-deviance}):
\begin{equation}
\mathbf{b} =  \frac{1}{2} \mathbf{diag}( {\bf R}_{xe}^\top \cdot {\bf R}^{-1}_{xx} \cdot {\bf R}_{xe} )/ \mathbf{diag}( {\bf R}_{ee} ),
\end{equation}
where the division between the two diagonal vectors is element-wise and 
$ {\bf R}_{xe} = {\bf R}_{xy} - {\bf R}_{xx} \cdot \hat{\bf H}$.
This bias term has to be computed for the full and reduced models, giving $\mathbf{b}_f$ and $\mathbf{b}_r$ respectively. The corresponding de-biased deviance is then: 
\begin{equation}
\mathscrbf{D}^{de-biased} = T  \log \left (\hat{\boldsymbol\sigma}^2_r / \hat{\boldsymbol\sigma}^2_f \right) - \mathbf{b}_r + \mathbf{b}_f,
\label{deviance-debiased}
\end{equation}
and can be used to compute the p-values as before. We have found empirically that we obtain a better (conservative) estimate of p-values if we use $T' = T - N$ instead of $T$ in this calculation of the de-biased deviance. $T'$ represents the effective degrees of freedom of the full model and converges to $T$ in the asymptotic limit for which the de-biased deviance formula was derived. 

\subsection{Validation of p-values with L2 regularization}
\label{section:validation-debiased-deviance}

\begin{figure}[ht]
\begin{center}
\includegraphics[width=0.7\columnwidth]{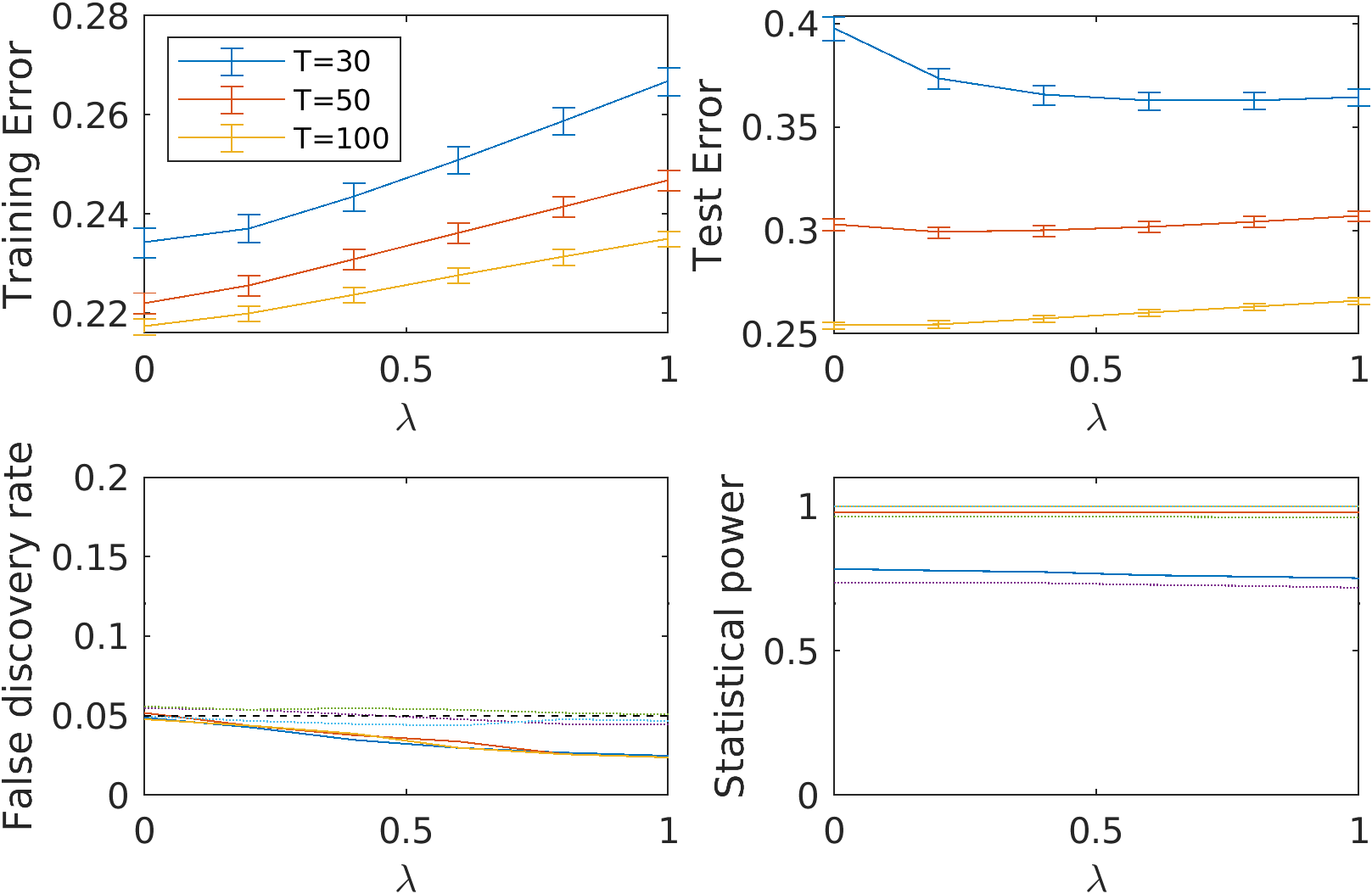}
\end{center}
\caption{Effect of L2 regularization on small toy examples $d_y=2, d_x=2$.   We ran the model and fit the data computing analysis of p-values in 1000 simulations with normal distributed error (zero mean, std of 2) and inputs x (zero mean, std of 1). (a) the training error is the relative error, i.e. $\sigma_e^2/std(y)^2$ computed on the same data that the model was fit on (length $T$ samples). Error bars indicate SEM over the 1000 simulations. (b) test error is the same relative error but computed on newly simulated data with the estimated model. (c) False discovery is the fraction of times where the null path is p<0.05. Solid lines are for false discovery in ${\bf A}$, and dotted lines are for false discovery in ${\bf B}$ in this simulation. (d) Statistical power is the fraction of times that the $p>0.05$ for the path that has non-zero coefficients. The specific models tested here had effects of $y_1 \rightarrow y_2$ and $x_2 \rightarrow y_2$ set to zero and otherwise $n_a=2, n_b=10$ coefficients with non-zero values.
}
\label{regularization}
\end{figure}

As $T$ increases the need for regularization decreases, so we scale the regularization factor as $\gamma = \lambda/ \sqrt{df}$, where $\lambda$ is a choice of regulation in the order of 1.  This particular scaling was established empirically to accomplish two things. One, if $\gamma$ is kept large for large $T$, then the asymptotic approximation used to compute the de-biased deviance is no longer correct and $p$-values are miss-estimated. Second, the specific scaling with the square root is an empirical finding that places the optimal $\gamma$ at similar values of $\lambda$  in simulation with different signal lengths $T$.  In Fig.~\ref{regularization} we simulated a simple model for varying $\lambda$ and $T$.  In this example we find, as expected, that training error increases with regularization (Fig.~\ref{regularization}a), but test error decreases with increasing regularization (Fig.~\ref{regularization}b). This effect becomes clear with larger noise (note that the explained variance here is relatively small) and with increasing correlation in input (here random normal inputs were drawn independently). The second observation is that with increasing sample size $T$, test error drops, and regularization becomes less important. Finally, we see that the bias correction provides conservative control of the false discovery rate (Fig.~\ref{regularization}c). The expected value in this example is $0.05$ and at small $T$ we are well below that. Indeed, for small $T$ we seem to be over-correcting, which limits the power to detect a true effect Fig.~\ref{regularization}d).

\subsection{Basis functions for the moving average filters}
\label{section:basis-functions}

The filter length (number of parameters) used in AR filters is typically kept relatively short, to avoid over-fitting, reduce the odds of instability in the recursion, and because even a single delay can already represent an infinite impulse response. This is not the case for MA filters, where longer responses have to be modeled explicitly, which can result in a relatively large number of parameters, with a risk of over-fitting. We have found empirically that the corrections we introduced in the Deviance estimate for short signals (Eq.~\ref{deviance-debiased}) do not work well when the filter lengths for ${\bf A}$ and ${\bf B}$ are very different. A solution to both these problems (imbalance in number of parameters and filter length) is to use basis functions for the ${\bf B}$ filters, following an approach used previously for TRFs \cite{Akram2017-ya, Miran2018-dx, Karunathilake2023-sy}. In this formalism, we have:
\begin{equation} 
{\bf B} = \underline{\bf B} \circ {\bf W},
\end{equation}
where the inner product $\circ$ is along the lag-axis of the filter matrix ${\bf B}$, and the goal now is to find the optimal $\underline{\bf B}$. The matrix ${\bf W}$ has dimensions $[n_b,\underline{n}]$ so that the number of parameters per filter is reduced from $n_b$ to $\underline{n}$. The linear least-squares problem remains unchanged with the closed-form solutions using now $\underline{\bf X} = {\bf W} {\bf X}$. In the equations above this can be implemented by replacing ${\bf R}_{xx}$ and ${\bf R}_{xy}$ with: 
\begin{eqnarray}
{\bf R}_{\underline{xx}} &=& {\bf W} \circ {\bf R}_{xx} \circ {\bf W}^\top, \\
{\bf R}_{\underline{x}y} &=& {\bf W} \circ {\bf R}_{xy}. 
\end{eqnarray}
Note that the new ${\bf R}_{\underline{xx}}$  and ${\bf R}_{\underline{x}y}$ are no longer Toeplitz matrices. The Granger formalism applies without change. 
  
\begin{figure}[ht]
\begin{center}
\includegraphics[width=1\columnwidth]{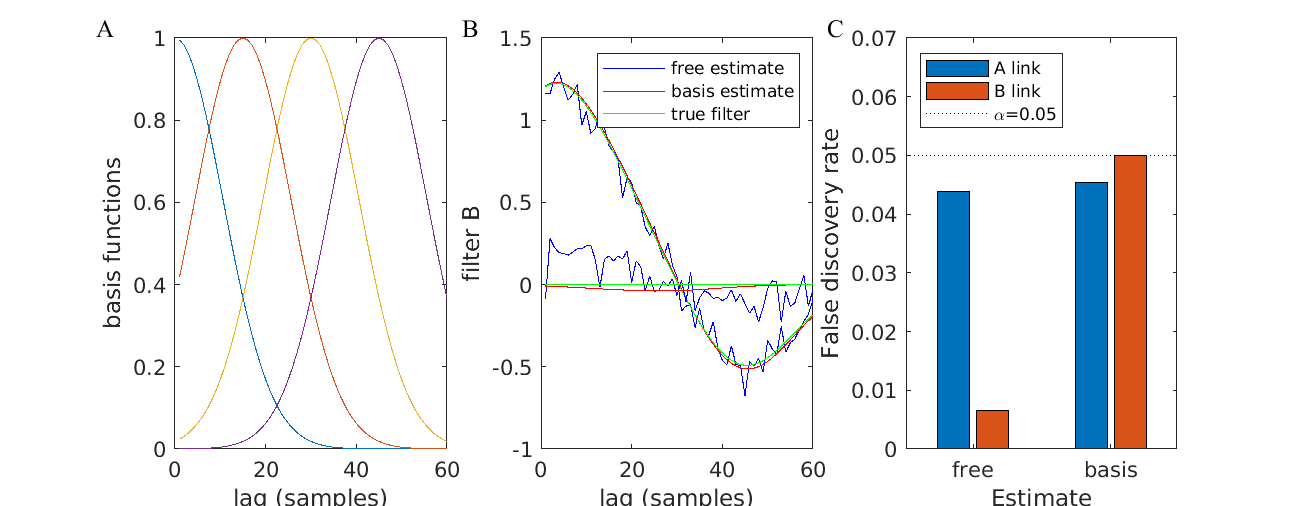}
\end{center}
\caption{Estimate MA filter coefficients ${\bf B}$ with and without basis functions (basis vs free, respectively). (A) The Gaussian basis functions used here, both for the generation of the data and for estimation. (B) Corresponding filter estimates and true filters used during data generation.  Here $d_y=2$ and $d_x=1$, with $y_1 \rightarrow y_2$ and $x \rightarrow y_2$ channels set to zero. Data was simulated with $T=300$ and $n_a=3$, $n_b=60$, $n_bar=4$. The total number of free parameters was $4*n_a+2*n_b=132$ (free) and $4*na+2*\underline{n}=20$ (basis). No regularization was used. (C) False discovery rate for AR and MA channels with and without basis functions. False discovery (bars) indicates how many times in 5000 random simulations these missing links reported a $p<0.05$, therefore we expect discovery rate to be 0.05 (dotted line).  
}
\label{basis}
\end{figure}

Here we implemented Gaussian windows as shown in Figure \ref{basis}A. With this, we are not only reducing the number of parameters, i.e. regularizing the solutions, but also selecting among a set of smooth filters ${\bf B}$. 

For large $T$ there is minimal regularization and regularization can not be used to “smooth” the estimated filters. For that, we developed the basis functions decomposition using smooth basis functions. A second motivation was also to make the number of parameters of ${\bf A}$ and ${\bf B}$ filters comparable, as we noticed empirically that an imbalance results in quite skewed $p$-values.  We simulated examples with very different filter lengths $n_a=2, n_b=60, n=6$ (Fig.~\ref{basis}). We find that introducing basis functions improves the estimates of false discovery rate while providing smooth filter estimates. We noticed empirically that these basis functions, for slow signals, made the signals significantly larger in power. As a result, conventional ridge regression regularized different signals differently, resulting in a miss-estimation of values. This motivated the adoption of Tikhonov regularization so that all variables are regularized by the same amount. 

\subsection{The case of a missing and superfluous variable}
\label{section:confound-collider}

\begin{figure}[ht]
\begin{center}
\includegraphics[width=0.6\columnwidth]{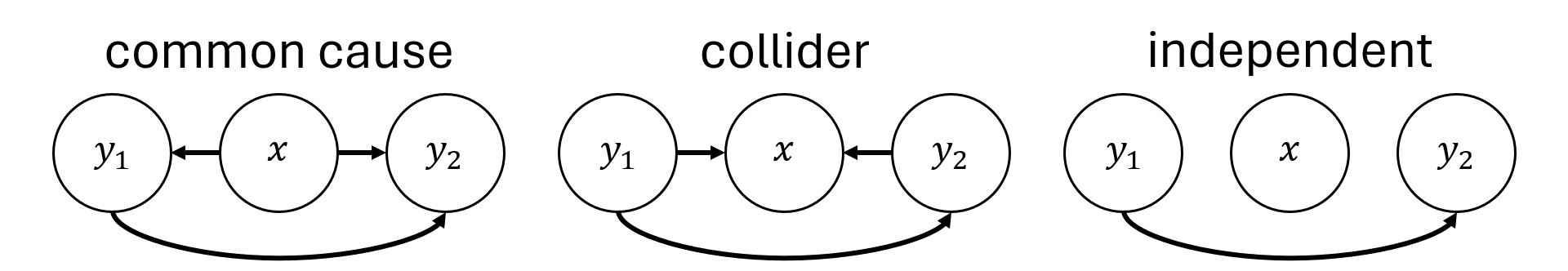}\\
\includegraphics[width=1\columnwidth]{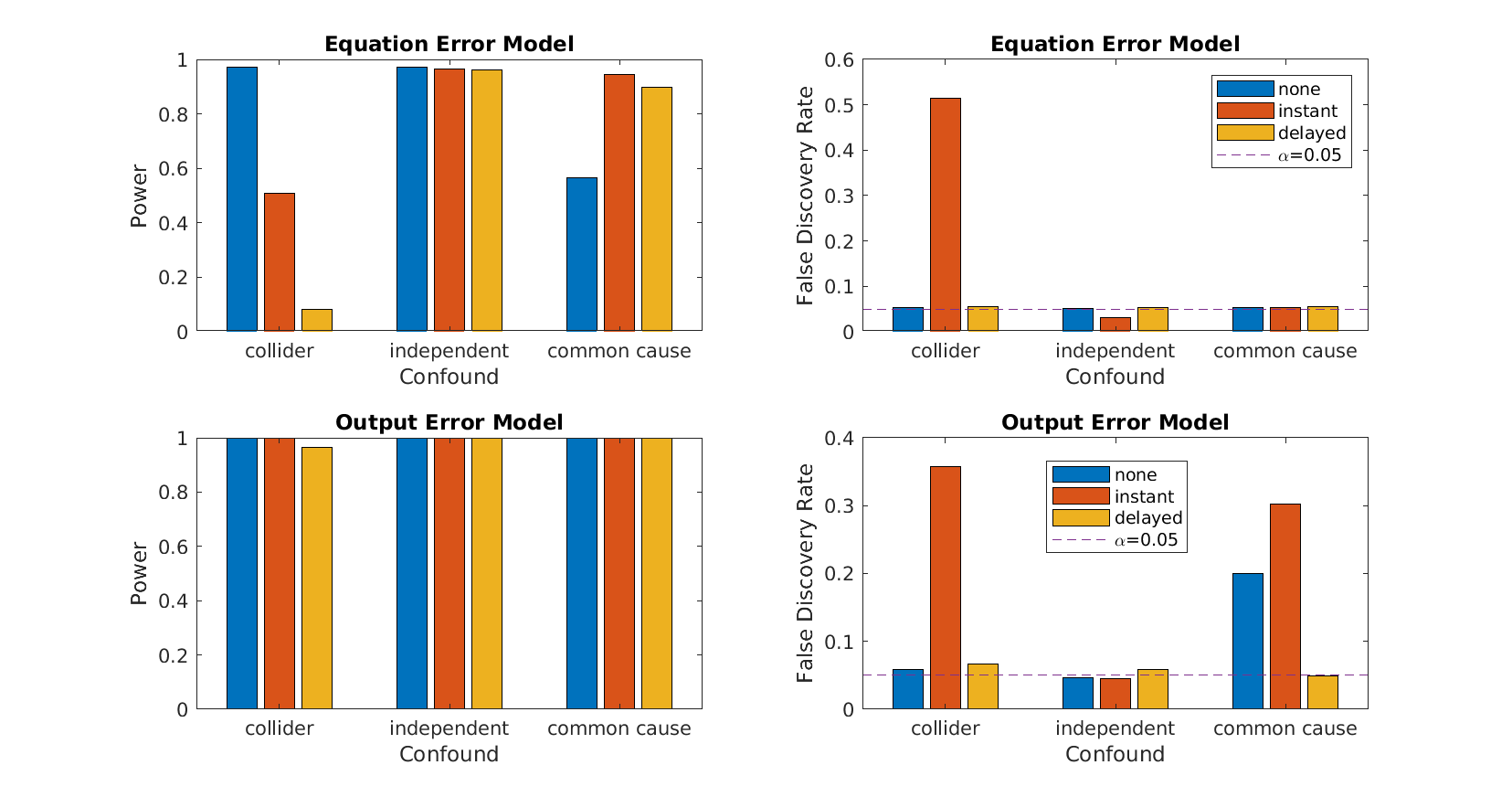}
\end{center}
\caption{The VARX (equation error) model (\ref{varx}) requires a larger $T$ to obtain similar power as the output error model (\ref{output-error}). Simulation here used $n_a=n_b=3, T=5000$ and normal i.i.d. error. }
\label{granger_simulation}
\end{figure}

Here we want to evaluate the case where the model does not match the data generation process. We simulated three possible data generation processes with one input and two outputs ($d_x=1, d_y=2$). We simulated three cases, where the exogenous input $x$ (conditionaing variable) will be either a common cause, a collider, or an independent variable (see Figure \ref{granger_simulation}). In all cases, the simulation implements a one-directional effect $y_1 \rightarrow y_2$. We then measure how frequently we find $p<0.05$ for this path, i.e. the power of the test to identify a correct path, and how frequently we find $p<0.05$ for $y_1 \leftarrow y_2$, i.e. the rate of false discovery. Note that conditioning on a collider is known to introduce spurious correlations \cite{Pearl2013-ce}. We test this for data generated with both the equation (VARX) and output error models. The results in Figure~\ref{granger_simulation} indicate the false discovery rate is correctly estimated at 0.05 in most scenarios, i.e. we are not finding causal effects above chance where there were none. This result holds regardless of whether $x$ was included as input (i.e. as a control variable with instant or delayed effect) or whether it did or did not have a true effect on $y_1$ and $y_2$ (common cause vs independent). Only when incorrectly modeling a collider as input, did we obtain spurious effects. Statistical power was improved when including the input to the model. In summary, there is no risk of false discovery when including input variables, even if they don't have a true effect, except if they are actually affected by the internal variables $y$. 

\subsection{Equation error versus output error model}
\label{section:output-error}

\begin{figure}[ht]
\begin{center}
\includegraphics[trim={9.5in 0 0 .7in},clip,width=.7\columnwidth]{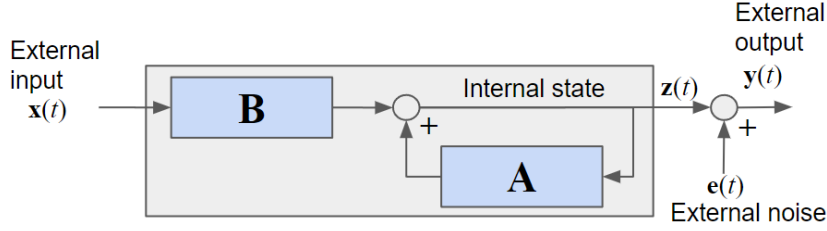}
\end{center}
\caption{Output error model: Here the dynamical variables ${\bf y}(t)$ are not observable. The gray box represents the overall system response $\mathcal{H}$.}
\label{schematic-output-error}
\end{figure}

An additional source of mismatch between the model and the data generation process is where error or the source of innovation originates. 
The VARX model in Eq. (\ref{varx}) is also called “equation error” model \cite{ljung1999system} because the error breaks the equality of the MA and AR terms. The equation error model assumes that ${\bf y}(t)$ is directly observable. It is different from an “output error” model where the recursion has no error, but the recursive signal ${\bf z}(t)$ is hidden and only observed with additive noise (see Fig.~\ref{schematic-output-error}):

\begin{align}
\begin{split}
{\bf z}(t) &= {\bf A}*{\bf z}(t-1) + {\bf B}*{\bf x}(t),\\  
{\bf y}(t) &= {\bf z}(t) + {\bf e}(t). 
\end{split}
\label{output-error}
\end{align}

For the output error model, the square error is not a quadratic function of the parameters ${\bf A}$ and ${\bf B}$. Therefore, there is no closed-form solution to the system identification problem, as we had for the equation error model. A few different iterative optimization approaches have been proposed, such as an expectation maximization (EM) algorithm \cite{Shumway1982-va}, gradient back-propagation through time \cite{Shynk1989-yt,igual2019adaptive}, or “pseudo regression” \cite{ljung1999system}. The pros and cons of the equation-error versus the output-error models are elaborated in the Discussion section. 

Continuing with the examples of the previous section (Fig.~\ref{granger_simulation}), despite using the VARX model, when the data was generated with an output error model, the false discovery remains limited at the target of 0.05. However, a common unobserved cause can generate spurious effects $y_1 \leftarrow y_2$. It has been suggested that running a Granger causality model on time-reversed data provides a control for this situation \cite{Vinck2015-ba, Winkler2016-ht}. We have found that running the model on time-reversed data results in spurious effects in all conditions tested here, so it is not clear to us how this can provide a remedy, and the caveat of an unobserved common cause remains when we are not directly observing endogenous variables, but only a noisy version of the internal dynamic.

\subsection{Derivation of the De-biased Deviance}
\label{section:derivation-debiased-deviance}

To our knowledge, the correction term for the deviance in the case of L2 regularization of a linear estimator is not yet available in the literature. We therefore detail he its derivation. 

Consider the linear model in Eq. (\ref{eq:linear-model}) for $n_y = 1$:
\begin{equation}
\mathbf{y} = \mathbf{X} \cdot \mathbf{h} + \mathbf{e},
\end{equation}
for simplicity of exposition. The case of $n_y > 1$ is a natural generalization, as we will explain below. A ridge regressor for $\mathbf{h}$ is obtained by:
\begin{equation}
\hat{\mathbf{h}} := \underset{\mathbf{h}}{\operatorname{argmin}} \quad \| \mathbf{y} -  \mathbf{X} \cdot \mathbf{h} \|^2 + \gamma \| \mathbf{h} \|^2,
\end{equation}
where $\gamma$ is the ridge parameter. To perform de-biasing, stationarity conditions for $\hat{\mathbf{h}}$ give:
\begin{equation}
-\mathbf{X}^\top (\mathbf{y} -  \mathbf{X} \hat{\mathbf{h}}) + \lambda \hat{\mathbf{h}} = \mathbf{0}.
\end{equation}
Following the arguments of Eqs. (3)--(5) in \cite{van2014asymptotically}, the following estimator is the de-biased estimator:
\begin{equation}
\hat{\mathbf{h}}^{de-biased} = \hat{\mathbf{h}} + (\mathbf{X}^\top \mathbf{X})^{-1}\mathbf{X}^\top \left(\mathbf{y} - \mathbf{X} \hat{\mathbf{h}}\right) = \hat{\mathbf{h}} + \mathbf{R}_{xx}^{-1} \mathbf{r}_{xe}.
\end{equation}
Now, consider the GC inference setting where we have full and reduce models given by:
\begin{equation}
\textrm{Full model:} \quad \mathbf{y} = \mathbf{X}^{ f} \mathbf{h}^{ f} + \mathbf{e}^{ f}, \quad \textrm{Reduced model:} \quad \mathbf{y} = \mathbf{X}^{ r} \mathbf{h}^{ r} + \mathbf{e}^{ r}.
\end{equation}
The log-likelihoods of the full and reduced models are given by:
\begin{equation}
\ell^{ f} := -\frac{T}{2} \ln (2\pi) -\frac{T}{2} \ln \sigma^2_{ f} - \frac{1}{2\sigma_{ f}^2} \left \| \mathbf{y} - \mathbf{X}^{ f} \mathbf{h}^{ f} \right \|^2,
\end{equation}
\begin{equation}
\ell^{ r} := -\frac{T}{2} \ln (2\pi) -\frac{T}{2} \ln \sigma^2_{ r} - \frac{1}{2\sigma_{ r}^2} \left \| \mathbf{y} - \mathbf{X}^{ r} \mathbf{h}^{ r} \right \|^2,
\end{equation}
and the deviance difference is given by $\mathscr{D} := 2 \big(\hat{\ell}^{ f} - \hat{\ell}^{ r} \big)$, where $\hat{\ell}^{ f}$ and $\hat{\ell}^{ r}$ are, respectively, the full and reduced log-likelihoods evaluated at the ridge estimates of full and reduced parameters:
\begin{align}
\hat{\mathbf{h}}^{f} &:= \underset{\mathbf{h}}{\operatorname{argmin}} \quad \| \mathbf{y} -  \mathbf{X}^{ f} \mathbf{h} \|^2 + \lambda \| \mathbf{h} \|^2,\\
\hat{\mathbf{h}}^{r} &:= \underset{\mathbf{h}}{\operatorname{argmin}} \quad \| \mathbf{y} -  \mathbf{X}^{ r} \mathbf{h} \|^2 + \lambda \| \mathbf{h} \|^2.
\end{align}
If the log-likelihoods were evaluated at the ML estimators, then $\mathscr{D}$ would be asymptotically chi-square/non-central chi-square, allowing to perform precise statistical tests. But, the ridge estimator is different than the ML estimator. So, we need to de-bias the deviance difference.

Following Eq. (20) and Eqs. (B.2)--(B.6) in \cite{Soleimani2022-bh}, the de-biased deviance difference is given by:
\begin{equation}
\mathscr{D}^{ de-biased} = 2 \big(\hat{\ell}^{ f} - \hat{\ell}^{ r} \big) - B_r + B_f,
\end{equation}
where the bias function $B(\cdot)$ is defined as:
\begin{align}
b_f &:=  \frac{1}{2 \hat{\sigma}^2_{ f}}\big(\mathbf{y} - \mathbf{X}^{ f} \hat{\mathbf{h}}^{f}\big)^\top \mathbf{X}^{ f} \left ( \mathbf{X}^{ f \top} \mathbf{X}^{ f} \right)^{-1} \mathbf{X}^{ f \top} \big(\mathbf{y} - \mathbf{X}^{ f} \hat{\mathbf{h}}^{f}\big) = \frac{1}{2 r^{f}_{ee}}\mathbf{r}_{xe}^{f \top} \mathbf{R}_{xx}^{f\, \, -1} \mathbf{r}^{f}_{xe},\\
b_r &:=  \frac{1}{2 \hat{\sigma}^2_{ r}}\big(\mathbf{y} - \mathbf{X}^{ r} \hat{\mathbf{h}}^{r}\big)^\top \mathbf{X}^{ r} \left ( \mathbf{X}^{ r \top} \mathbf{X}^{ r} \right)^{-1} \mathbf{X}^{ r \top} \big(\mathbf{y} - \mathbf{X}^{ r} \hat{\mathbf{h}}^{r}\big) = \frac{1}{2 r^{r}_{ee}}\mathbf{r}_{xe}^{r \top} \mathbf{R}_{xx}^{r\, \, -1} \mathbf{r}^{r}_{xe}.
\end{align}
Simplifying the deviance difference $\mathscr{D}$ for the linear model, we can express $\mathscr{D}^{de-biased}$ as:
\begin{equation}
\mathscr{D}^{de-biased} = T \ln \frac{\hat{\sigma}^2_{ r}}{\hat{\sigma}^2_{ f}} - b_r + b_f,
\end{equation}
where 
\begin{equation}
\hat{\sigma}_{ f}^2 := \frac{1}{T} \left \| \mathbf{y} - \mathbf{X}^{ f} \hat{\mathbf{h}}^{f} \right\|^2 = r_{ee}^{f}, \qquad \hat{\sigma}_{ r}^2 := \frac{1}{T} \left \| \mathbf{y} - \mathbf{X}^{ r} \hat{\mathbf{h}}^{r} \right\|^2 = r_{ee}^{r},
\end{equation}
are the full and reduced prediction variances. For $n_y > 1$, given that each channel of $\mathbf{Y}$ is treated separately, the full and reduced bias vectors are given by:
\begin{equation}
\mathbf{b}_f =  \frac{1}{2} \mathbf{diag}( {\bf R}_{xe}^{f \top} \cdot {\bf R}^{f \, \, -1}_{xx} \cdot {\bf R}_{xe}^f )/ \mathbf{diag}( {\bf R}_{ee}^{f} ), \qquad \mathbf{b}_r =  \frac{1}{2} \mathbf{diag}( {\bf R}_{xe}^{r \top} \cdot {\bf R}^{r \, \, -1}_{xx} \cdot {\bf R}_{xe}^r )/ \mathbf{diag}( {\bf R}_{ee}^{r} ),
\end{equation}
and the de-biased deviance vector is given by:
\begin{equation}
\mathscrbf{D}^{de-biased} = T  \log \left (\hat{\boldsymbol\sigma}^2_r / \hat{\boldsymbol\sigma}^2_f \right) - \mathbf{b}_r + \mathbf{b}_f.
\end{equation}

Under the null hypothesis, the de-biased deviance difference can be shown to be asymptotically, as $T \rightarrow \infty$, chi-square distributed with $n$ degrees of freedom, where $n$ is the number of parameters removed in the reduced model and $T$ is the number of observations (i.e., column dimension of $\mathbf{Y}$) \cite{Davidson1961-dr}. This asymptotic property requires that ridge estimator is consistent. To guarantee this, $\gamma$ must grow slower than $\mathcal{O}(T)$. In practice, fixing $\gamma$ using cross-validation typically meets this condition.
\bibliographystyle{plos2015} 
\bibliography{references}

\end{document}